\documentclass{article}
\usepackage{ijcai21}

\usepackage{times}
\usepackage{soul}
\usepackage{url}
\usepackage[hidelinks]{hyperref}
\usepackage[utf8]{inputenc}
\usepackage[small]{caption}
\usepackage{graphicx}
\usepackage{amsmath}
\usepackage{amsthm}
\usepackage{booktabs}
\usepackage{algorithm}
\usepackage{algorithmic}
\usepackage{enumitem}
\title{FedSpeech: Federated Text-to-Speech with Continual Learning}

\author{
Ziyue Jiang$^1$\and
Yi Ren$^1$\and
Ming Lei$^{2}$\And
Zhou Zhao$^1$\\
\affiliations
$^1$Zhejiang University\\
$^2$Alibaba Group\\
\emails
ziyuejiang341@gmail.com,
rayeren@zju.edu.cn,
lm86501@alibaba-inc.com,
zhaozhou@zju.edu.cn
}

\begin{document}

\maketitle

\begin{figure*}[htbp]
    \centering
    \begin{minipage}[t]{1.0\textwidth}
        \centering
        \includegraphics[width=16cm, clip=true]{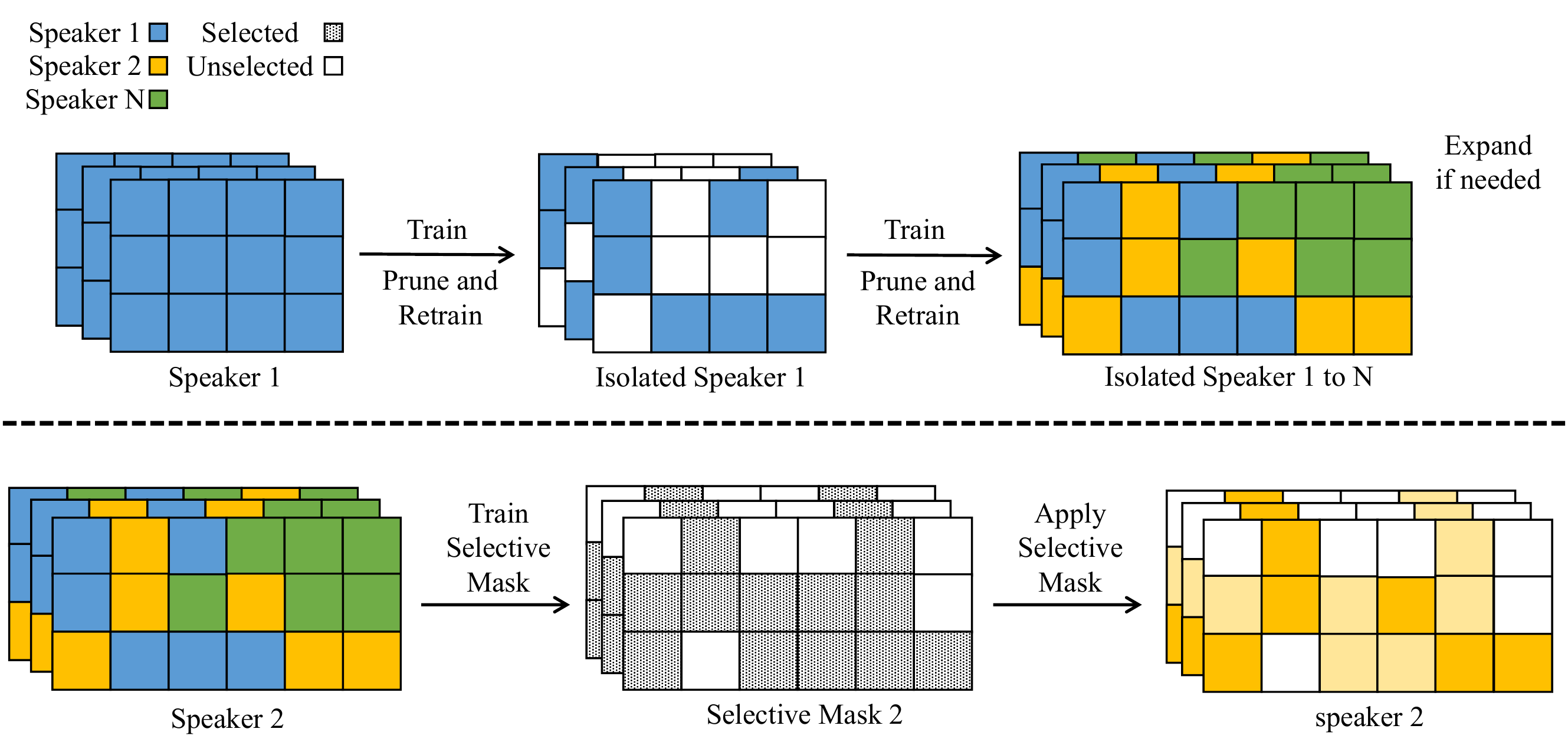}
    \end{minipage}
    \caption{ The two rounds of training process with FedSpeech. In Round 1, gradual pruning masks are performed to isolate weights for each speaker. If the weights preserved for a certain speaker are less than the threshold, the model will expand. In Round 2, we take speaker 2 as the example. Selective masks are trained to reuse the knowledge from the weights preserved for other speakers.}
    \vspace{-0.2cm}
	\label{fig_two_round}
\end{figure*}

\begin{abstract}
Federated learning enables collaborative training of machine learning models under strict privacy restrictions and federated text-to-speech aims to synthesize natural speech of multiple users with a few audio training samples stored in their devices locally. However, federated text-to-speech faces several challenges: very few training samples from each speaker are available, training samples are all stored in local device of each user, and global model is vulnerable to various attacks. In this paper, we propose a novel federated learning architecture based on continual learning approaches to overcome the difficulties above. Specifically, 1) we use gradual pruning masks to isolate parameters for preserving speakers' tones; 2) we apply selective masks for effectively reusing knowledge from tasks; 3) a private speaker embedding is introduced to keep users' privacy. Experiments on a reduced VCTK dataset demonstrate the effectiveness of FedSpeech: it nearly matches multi-task training in terms of multi-speaker speech quality; moreover, it sufficiently retains the speakers' tones and even outperforms the multi-task training in the speaker similarity experiment.

\end{abstract}

\section{Introduction}
Federated learning has become an extremely hot research topic due to the ``data island" issue in recent years. After the concept of federated learning is proposed by Google recently ~\cite{konevcny2016federated}, most federated learning researches focus on computer vision ~\cite{mcmahan2017communication}, natural language processing ~\cite{hard2018federated}, and voice recognition ~\cite{leroy2019federated}. Collaborative training brings great improvements to the performance of those tasks. However, few of them pay attention to federated text-to-speech (TTS) tasks which aim to synthesize natural speech of multiple users with a few audio training samples stored in each user's local device. Due to the nature of current neural network based systems, federated TTS tasks face several challenges below:
\begin{itemize}[leftmargin=*]

\item \textbf{Lack of training data.} Although neural network-based TTS models ~\cite{shen2018natural,ren2020fastspeech} can generate high-quality speech samples, they suffer from the lack of robustness (e.g., word skipping and noise) and significant quality degradation when the training dataset's size is reduced. However, in federated TTS scenarios, each user has a limited number of audio training samples, especially for low-resource languages.

\item \textbf{Strict data privacy restrictions.} In federated scenarios, training samples are all stored in device of each user locally, which makes multi-task (multi-speaker) training impossible. However, in typical federated aggregation training ~\cite{mcmahan2017communication}, even small gradients from other speakers may greatly hurt the tone of a specific speaker due to the catastrophic forgetting issues. Accurately maintaining the tone of each speaker is difficult.

\item \textbf{Global model is vulnerable to various attacks.} Typical communication architectures used in federated learning ~\cite{mcmahan2017communication,rothchild2020fetchsgd} aggregate the information (e.g., gradients or model parameters) and train a global model. Although the local data are not exposed, the global model may leak sensitive information about users.

\end{itemize}

However, traditional federated learning methods can not address the issues above simultaneously. Recently, continual lifelong learning has received much attention in deep learning studies. Among them, Hung \textit{et al.} have proposed a method called "Compacting, Picking and Growing" ~\cite{hung2019compacting}, which uses masks to overcome catastrophic forgetting in continual learning scenarios and has achieved significant improvements in image classification tasks. Inspired by their works, we focus on building a federated TTS system using continual learning techniques.

Thus, in order to bring the advantages of collaborative training into federated multi-speaker TTS systems, in this work, we propose a federated TTS architecture called FedSpeech, in which 1) we use gradual pruning masks to isolate parameters for preserving speakers' tones; 2) we apply selective masks for effectively reusing the knowledge from tasks under privacy restrictions; 3) a private speaker embedding is introduced to apply additional information and guarantee users' privacy. Different from Hung \textit{et al.} \shortcite{hung2019compacting}, we perform masks in a Transformer-based TTS model and select weights from both previous and later tasks to make it more equitable for all speakers. Our proposed FedSpeech can address the above-mentioned three challenges as follows:
\begin{itemize}[leftmargin=*]

\item With selective masks, FedSpeech can effectively benefit from collaborative training to lessen the influence of limited training data.

\item Gradual pruning masks isolate the parameters of different speakers to overcome catastrophic forgetting issues. Thus, FedSpeech avoids the issue of tone changes for all speakers.

\item The private speaker embedding is introduced coupled with two types of masks above to preserve the privacy and avoid various attacks for speakers.
\end{itemize}

We conduct experiments on a reduced VCTK dataset\footnote{In order to simulate the low-resource language scenarios, we randomly select 100 audio samples from each speaker for training.} to evaluate FedSpeech. The results show that in terms of speech quality, FedSpeech is nearly equivalent to joint training, which breaks the privacy rules. Moreover, FedSpeech achieves even higher speaker similarity scores than joint training due to our parameter isolation policy. We attach some audio files generated by our method in the supplementary materials\footnote{Synthesized speech samples can be found in: \url{https://fedspeech.github.io/FedSpeech_example/}}.

\section{Background}
In this section, we briefly overview the background of this work, including text to speech (TTS), federated learning, and continual learning.

\paragraph{Text to Speech.} \label{bg_tts}
Aiming to synthesize natural speech, text to speech (TTS)~\cite{shen2018natural,yamamoto2020parallel} remains a crucial research topic. Recently, many advanced techniques have dominated this field. From concatenative synthesis with unit selection, statistical parametric synthesis to neural network based parametric synthesis and end-to-end models~\cite{shen2018natural,ren2019fastspeech,ren2020fastspeech,DBLP:conf/nips/KimKKY20}, the quality of the synthesized speech is closer to the human voice. However, in federated TTS tasks, most neural TTS systems generate audios with quality decline when the training dataset's size is significantly reduced. In this work, we efficiently reuse knowledge from different tasks, which resolves the problem above.

\paragraph{Federated Learning.} \label{bg_tts}
Strictly under the privacy restrictions among distributed edge devices, federated learning aims to train machine learning models collaboratively~\cite{li2019survey}. In federated learning, typical communication architectures~\cite{mcmahan2017communication,hard2018federated,rothchild2020fetchsgd} usually aggregate the information (e.g., gradients or model parameters) and train a global model, which can be centralized design or decentralized design. However, the proposed methods above are vulnerable to inference attacks and may expose sensitive information about the users in TTS tasks. Moreover, the gradients from the other speakers may result in catastrophic forgetting and greatly hurt one speaker's tone. In this work, we utilize a private speaker embedding policy to protect users' privacy. Besides, we adopt two kinds of parameter masks in the training process and combine them in the inference process to retain tones and transfer knowledge.

\paragraph{Continual Learning.} \label{bg_tts}
Continual learning aims at overcoming the catastrophic forgetting issues of neural networks when tasks arrive sequentially. In continual learning scenarios, a mechanism should be introduced to continually accumulate knowledge over different tasks without the need to retrain from scratch. Simultaneously, the mechanism should ensure a good compromise between the model's stability and plasticity. Continual learning approaches are mainly structured into three main groups: replay, regularization-based, and parameter isolation methods. Replay methods~\cite{rolnick2019experience} explicitly retrain on a subset of stored samples from previous tasks, which breaks the users' privacy rules. Prioritizing privacy, regularization-based methods~\cite{li2017learning} introduce new loss functions to distill previous knowledge or penalize important parameters' updates. However, slight weight changes may significantly affect the voice of a specific speaker. Considering the privacy and performance, the parameter isolation method~\cite{mallya2018packnet,hung2019compacting} is a solution to the difficulties above. Our FedSpeech adopts the progressive pruning masks in Hung \textit{et al.} \shortcite{hung2019compacting} and modifies the selective masks referring to Mallya \textit{et al.} \shortcite{mallya2018piggyback} to transfer knowledge from other speakers while keeping privacy. 

\begin{figure}[t!]
\centering
\includegraphics[width=0.46\textwidth]{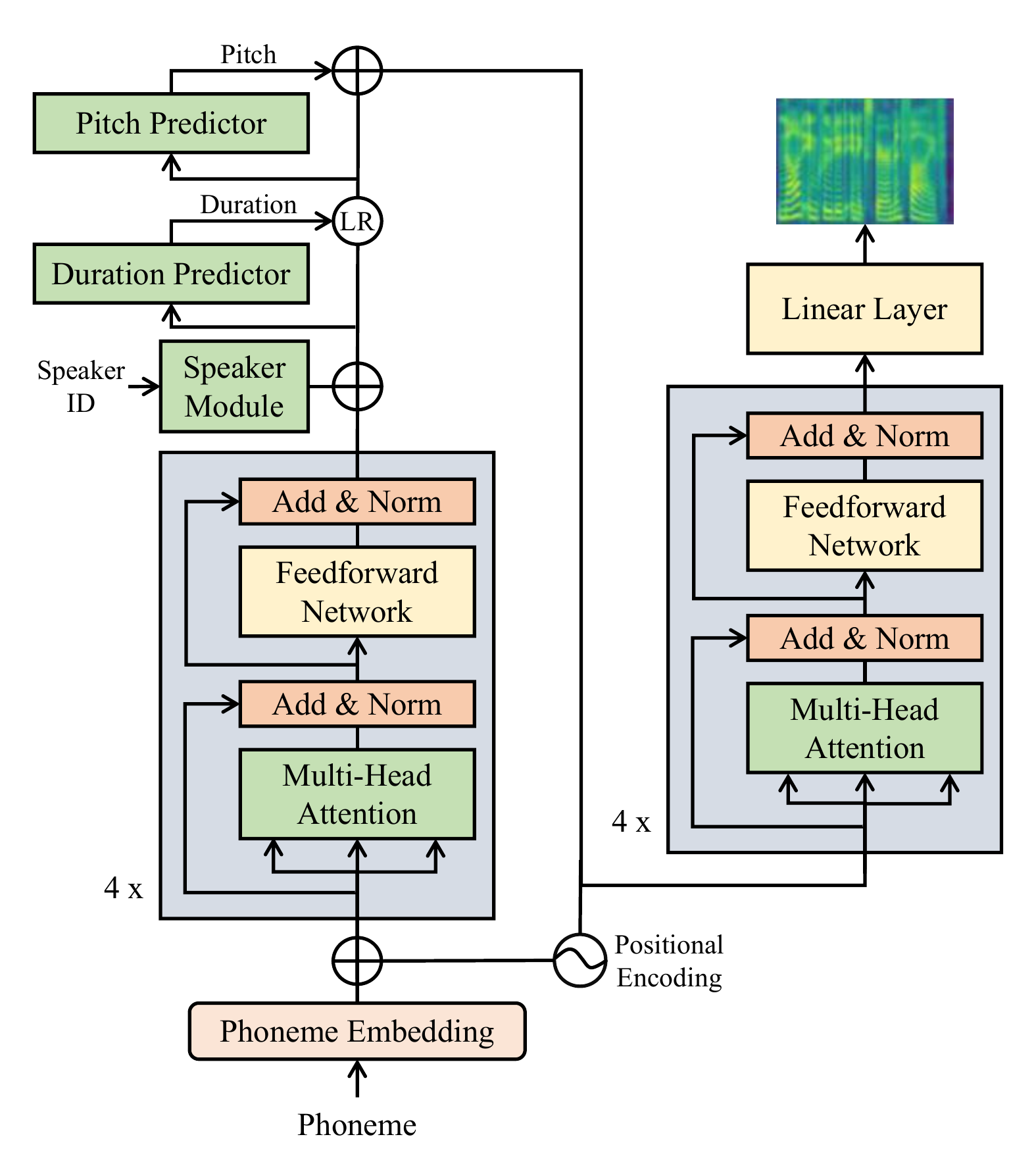}
\caption{The overall architecture for FedSpeech. \textcircled{$+$} denotes the element-wise addition operation. LR denotes the length regulator proposed in FastSpeech \protect\cite{ren2019fastspeech}. }
\label{fig:fig_fs2}
\end{figure}

\section{FedSpeech}
In this section, we first introduce the architecture design of FedSpeech based on the novel feed-forward transformer structure proposed in Ren \textit{et al.} \shortcite{ren2020fastspeech}. Then we introduce a private speaker embedding to apply additional information and keep users' privacy. It is trained to capture the speaker features from the latent space and stores sensitive information, which is indispensable in the inference stage. The private speaker embedding should be preserved in each user's device locally and can not be obtained by other users in order to protect users’ privacy. In order to solve the data scarcity issue and further address the privacy issue in federated TTS tasks, we propose a two-round sequential training process, which is a common setting in continual learning. We adopt two kinds of masks and a speaker module to isolate parameters and effectively reuse different speakers' knowledge while keeping privacy. We describe the above components in detail in the following subsections.

\subsection{FedSpeech Architecture}
The overall model architecture of FedSpeech is shown in Figure \ref{fig:fig_fs2}. The encoder converts the phoneme embedding sequence into the phoneme hidden sequence, and then we add different variance information such as duration and pitch into the hidden sequence, finally the mel-spectrogram decoder converts the adapted hidden sequence into mel-spectrogram sequence in parallel. We adopt the feed-forward Transformer block, which is a stack of self-attention \cite{vaswani2017attention} layer and 1D-convolution feed-forward network as in FastSpeech ~\cite{ren2019fastspeech}, as the basic structure for the encoder and mel-spectrogram decoder. Besides, we adopt a pitch predictor and a duration predictor to introduce more information. Each of them consists of a 2-layer 1D-convolutional network with ReLU activation, followed by the layer normalization and the dropout layer, and an extra linear layer to project the hidden states into the output sequence. In training, we take the ground-truth values of duration and pitch extracted from the recordings as input into the hidden sequence to predict the target speech. Meanwhile, we use the ground-truth duration and pitch values as targets to train the predictors. And their outputs are used in inference to synthesize target speeches.

\subsection{Speaker Module}
In order to control voice by estimating the speaker features from the latent space and protect privacy, we introduce a private speaker module which is a trainable lookup table \cite{jia2018transfer} taking speaker's identity number $S_{id}$ as input and generating the speaker representation $R = \{r_1,r_2,...,r_n\}$, where $n$ is the hidden size of the model. And then the speaker representation $R$ is passed to the output of the encoder as additional key information to control the tone characteristics in both training and inference. Each speaker trains and keeps his own set of module parameters in consideration of privacy so that the others can not synthesize his voice even with his $S_{id}$.

\subsection{Two Rounds of Sequential Training}
Our work abandons the federated aggregation training setup, due to its catastrophic forgetting issue. As shown in Figure \ref{fig_two_round}, we follow a sequential training setup, which is common in continual learning \cite{mallya2018packnet,mallya2018piggyback,hung2019compacting}. It is worth noting that only knowledge from previous speakers can be used by the current speaker in Hung \textit{et al.} \shortcite{hung2019compacting}. Different from their works, we propose the two rounds of sequential training. In the first round of training, the model separately learns and fixes a portion of weights for each speaker, so that in the second round of training we can selectively reuse the knowledge from both previous and later speakers. In the following, we present our two training rounds in detail.

\subsubsection{First Round - Gradual Pruning Masks}
\label{sec_gradual_pruning_masks}
In the first round of training, the gradual pruning masks in Figure \ref{fig_two_round} are calculated to isolate parameters for each speaker. The speakers from $1$ to $N$ are denoted as $S_{1:N}$. The tasks for $S_{1:N}$ are $T_{1:N}$ and start sequentially. For simplicity, we take $S_{t}$ as the example. When $T_{t}$ starts, the global model $M_{g}$ is firstly sent to $S_t$ and trained using his private data until convergence. The learned weight matrix for layer $i$ is denoted as $W_{1}^{l_i}$. We then gradually prune a portion of the smallest weights in $W_{1}^{l_i}$ for each layer, set them to $0$, and retrain the other weights to restore performance. Finally, the weights are divided into three parts: 1) the zero-valued weights released for later speakers $S_{t+1:N}$; 2) the fixed weights $W_{S}^{1:t-1}$ preserved by previous speakers $S_{1:t-1}$; 3) the weights $W_{S}^{t}$ preserved by $S_t$. If the released weights for later speakers $S_{t+1:N}$ are less than a threshold $\lambda$, we will expand the hidden size of the model by $\mu$. The pruning state is stored in the gradual pruning masks denoted as $m_p$. We then fix $W_{S}^{t}$ and send $m_p$ and $M_{g}$ (except for the private speaker module) to the device of next speaker $S_{t+1}$ to continue sequential training. When the first round ends, each speaker preserves a certain portion of weights denoted as $W_{S}^{1:N}$ represented by $m_p$. As the weights of each task are fixed, each speaker can perfectly retain their tones in inference. Finally, the $m_p$ and $M_{g}$ are sent to the devices of $S_{1:N}$. Thus, each speaker has $m_p$, $M_{g}$, and his preserved parameters of the private speaker module.

\begin{algorithm}[tb]
\caption{Two Rounds of Training with FedSpeech}
\label{alg:algorithm}
\textbf{Input}: number of tasks (speakers) $N$, training samples with paired text and audio $\mathcal{D}_t = \{(x_k, y_k)\}_{k=1}^K$ of speaker $t$. \\
\textbf{Initialize}: randomly initialize the original model $M_g$; set all elements in the gradually pruning masks $m_p$ to zeros; set all elements in the real-valued selective mask $m_s^t$ of speaker $t$ to $\alpha$ and the binary selective mask $m_b^t$ of speaker $t$ to zeros; initialize the threshold $\lambda$, $\mu$, and $\sigma$.  \\
\textbf{Annotate}: $T_{1:N}$ denote tasks from $1$ to $N$; $S_{1:N}$ denote speakers from $1$ to $N$; $W_S^{t}$ denotes the weights in $M_g$ preserved for speaker $t$.\\
\textbf{First Round}:
\begin{algorithmic}[0] 
\FOR{task $t=T_{1:N}$}
\STATE Train the released weights until convergence
\IF {$t \neq T_N$}
\STATE Gradually prune a portion of the smallest weights, set them to $0$ and retrain other weights to restore performance
\STATE Release the zero-valued weights for next task
\ENDIF
\STATE store the pruning state in $m_p$ and fix $W_{S}^{t}$
\IF {the released weights are less than $\lambda$}
\STATE Expand the hidden size of $M_g$ by $\mu$
\ENDIF
\ENDFOR
\STATE Send $m_p$ and $M_g$ to the local devices of $S_{1:N}$
\end{algorithmic}
\textbf{Second Round}:
\begin{algorithmic}[0] 
\STATE fix $W_{S}^{1:N}$
\FOR{task $t=T_{1:N}$ \textbf{in parallel in local devices}}
\STATE Initialize $m_s^t$
\STATE Train $m_s^t$, $m_b^t$ until convergence
\STATE $S_t$ preserves $m_b^t$ for inference
\ENDFOR
\end{algorithmic}
\end{algorithm}

\subsubsection{Second Round - Selective Masks}
\label{sec_selective_masks}
In the second round of training, we introduce the selective masks to transfer knowledge from speakers to address the data scarcity issue. The selective masks in Figure \ref{fig_two_round} are trained to automatically select useful weights preserved by speakers. Instead of selecting weights from previous tasks ~\cite{mallya2018piggyback}, we propose a modified selection procedure to select weights from all tasks, which is more equitable for every speaker (especially for more previous speakers) in federated TTS tasks. For a specific speaker $S_t$, 
our two rounds of training abandon the joint training of $W_S^t$ and the selective masks, which leads to slight performance degradation. But for each speaker, we make it possible to select weights from both previous and later tasks, which leads to significant improvements overall.

Assume that when the first round ends, the weights of $M_g$ are divided into several portions $W_{S}^{1:N}$ which are preserved by $S_{1:N}$. In order to benefit from collaborative training while keeping privacy, we introduce a learnable mask $m_b \in \{0,1\}$ to transfer knowledge from the parameters preserved by other speakers. We use the piggyback approach \cite{mallya2018piggyback} that learns a real-valued mask $m_s$ and applies a threshold for binarization to construct $m_b$. For a certain speaker $S_t$, mask $m_b^t$ is trained on his local dataset to pick the weights from other speakers by $m_b^t \odot W_{S}^{1:t-1} \cup W_{S}^{t+1:N}$.

We describe the mask training process of the selective masks in 1D convolution layer as the example. At task $t$, $M_{g}$ (i.e., $W_{S}^{1:N}$) is fixed. Denote the binary masks as $m_b^t$. Then the equation for the input-output relationship is given by
\begin{equation}
\tilde{W} = m_b^t \odot W ,
\end{equation}
\vspace{-0.45cm}
\begin{equation}
\resizebox{.91\linewidth}{!}{$
    \displaystyle
    y(N_i,C_{out_j}) = b(C_{out_j}) + \sum_{k=0}^{C_{in}-1} \tilde{W}(C_{out_j},k) \star x({N_i},k) ,
$}
\end{equation}
where $\star$ is the valid cross-correlation operator, $N$ is the batch size, and $C_{in}$/$C_{out}$ denote the number of in/output channels. 

In the backpropagation process, the $m_b^t$ is not differentiable. So we introduce the real-valued selective masks denoted as $m_s^t$. Denote $\sigma$ as the threshold for selection. As in Hung \textit{et al.} \shortcite{hung2019compacting}, when training the binary mask $m_b^t$, we update the real-valued mask $m_s^t$ in the backward pass; then $m_b^t$ is quantized with a binarizer function $\beta$ applied on $m_s^t$ and used in the forward pass. After training, we discard $m_s^t$ and only store $m_b^t$ for inference. The equation for $m_s^t$ is formulated by
\begin{equation}
m_b^t = \beta(m_s^t) = \left\{\begin{matrix}
                             1 & {\rm if}\;m_s^t > \sigma \\ 
                             0 & else 
                            \end{matrix}\right. ,
\label{equation_3}
\end{equation}
\begin{equation}
\resizebox{.91\linewidth}{!}{$
    \displaystyle
	\begin{split}
    \delta{m_s^t(C_{out_j},k)} &= \frac{\partial{L}}{\partial{m_b^t(C_{out_j},k)}} \\ 
    &= \frac{\partial{L}}{\partial{y(N_i,C_{out_j})}} \cdot \frac{\partial{y(N_i,C_{out_j})}}{\partial{m_b^t(C_{out_j},k)}} \\ 
    &=\delta{y(N_i,C_{out_j})} \cdot \sum_{k=0}^{C_{in}-1} \tilde{W}(C_{out_j},k) \star x({N_i},k)
	\end{split} .
	$}
\label{equation_4}
\end{equation}
With the Equation \eqref{equation_3} and \eqref{equation_4}, we are able to train the selective masks so as to select useful weights for each speaker.

\subsubsection{Model Inference}
For simplicity, we describe the inference stage using the example of $S_t$. Now $S_t$ has $m_p$, $m_b^t$, $M_g$ and his locally preserved parameters of speaker module.  We pick the weights $W_S^t$ using $m_p$ and selectively reuse the weights in $W_{S}^{1:t-1} \cup W_{S}^{t+1:N}$ using $m_b^t$. The unused weights are fixed to zero in order not to hurt the tone of $S_t$. The overall procedure of the two rounds of training with FedSpeech is presented in Algorithm \ref{alg:algorithm}.

\section{Experimental Setup}
\subsection{Datasets}
We conduct experiments on the VCTK dataset \cite{veaux2017superseded}, which contains approximate 44 hours of speech uttered by 109 native English speakers with various accents. Each speaker reads out about 400 sentences, most of which are selected from a newspaper plus the Rainbow Passage and an elicitation paragraph intended to identify the speaker's accent. To simulate the low-resource language scenarios, we randomly select and split the samples from each speaker into 3 sets: 100 samples for training, 20 samples for validation, and 20 samples for testing. We randomly select 10 speakers denoted as task 1 to 10 for evaluation respectively. To alleviate the mispronunciation problem, we convert the text sequence into the phoneme sequence with an open-source grapheme-to-phoneme conversion tool \cite{g2pE2019}. Following Shen \textit{et al.} \shortcite{shen2018natural}, we convert the raw waveform into mel-spectrograms and set frame size and hop size to 1024 and 256 with respect to the sample rate 22050. 

\subsection{Model Configuration}

\paragraph{FastSpeech 2 Model.} FedSpeech is based on FastSpeech 2 \cite{ren2020fastspeech}, which consists of 4 feed-forward Transformer blocks both in the encoder and the mel-spectrogram decoder. The hidden sizes of the self-attention and 1D convolution in each feed-forward Transformer block are all set to 256, which will grow if needed. The number of attention heads is set to 2. The output linear layer converts the 256-dimensional hidden into the 80-dimensional mel spectrogram. 

\paragraph{Gradual Pruning and Selective Masks.} In the training stage, the type of gradual pruning masks is short integer. We use speakers' identity number stored in the gradual pruning masks to mark the parameters so as to isolate them. The type of selective masks is parameter when training. After training, the selective masks will be quantized to binary type. The initial value of selective masks' unit is 0.01 and the $\sigma$ in the Equation \ref{equation_3} is set to 0.005.

\subsection{Training and Inference}
We use 1 Nvidia 1080 Ti GPU, with 11GB memory. Each batch contains about 20,000 mel-spectogram frames. We use Adam optimizer with $\beta_1 = 0.9$, $\beta_2 = 0.98$, $\epsilon = 10^{-9}$ and follow the learning rate schedule in \cite{vaswani2017attention}. In all experiments, we choose 10 speakers. For each speaker, it takes 4k steps for FedSpeech model training (including the gradual pruning masks' training) and 1k steps for selective masks' training. The total training time of FedSpeech for 10 speakers is 14 hours. For those baselines without these masks, we apply their approach to FastSpeech 2 model \cite{ren2020fastspeech} and train the model for 5k steps for a fair comparison. In the inference stage, we use the pre-trained Parallel WaveGAN (PWG) \cite{yamamoto2020parallel} to transform mel-spectrograms generated by FedSpeech into audio samples.

\begin{table}
\centering
\begin{tabular}{lrrrrr}  
\toprule
Method & MOS & $\gamma$ \\
\midrule
GT             & 3.97 $\pm$ 0.050 & - \\
GT (Mel + PWG) & 3.85 $\pm$ 0.051 & - \\
Multi-task  & 3.82 $\pm$ 0.050 & 1.7  \\
\midrule
Scratch     & 3.71 $\pm$ 0.056 & 10   \\
Finetune    & 3.72 $\pm$ 0.051 & 10   \\
FedAvg \cite{mcmahan2017communication}      & 3.58 $\pm$ 0.067 & 1.7  \\
CPG    \cite{hung2019compacting}            & 3.74 $\pm$ 0.053 & 1.7  \\
\midrule
FedSpeech   & 3.77 $\pm$ 0.052 & 1.7  \\
\bottomrule
\end{tabular}
\caption{The MOS with 95\% confidence intervals. $\gamma$ means the model expansion rate compared with FedSpeech with 256 hidden size.}
\label{table_1}
\end{table}

\section{Results}
In this section, we evaluate the performance of FedSpeech in terms of audio quality, speaker similarity, and ablation studies.

\subsection{Audio Quality}
\label{audio_quality}
We evaluate the MOS (mean opinion score) on the test set to measure the audio quality. The setting and the text content are consistent among different models so as to exclude other interference factors, only examining the audio quality. Each audio is judged by 10 native English speakers. We compare the MOS of the audio samples generated by our model with other systems, which include 1) \textit{GT}, the ground truth audio in VCTK. 2) \textit{GT (Mel + PWG)}, where we first convert the ground-truth audio into mel-spectrograms, and then convert the mel-spectrograms back to audio using Parallel WaveGAN (PWG) \cite{yamamoto2020parallel}; 3) \textit{Multi-task}, jointly training without privacy restrictions; 4) \textit{Scratch}, learning each task independently from scratch; 5) \textit{Finetune}, fine-tuning from a previous model randomly selected and repeats the process 5 times (for task 1, Finetune is equal to Scratch); 6) \textit{FedAvg} \cite{mcmahan2017communication}, aggregating the local information (e.g., gradients or model parameters) and train a global model. 7) \textit{CPG} \cite{hung2019compacting}, a parameter isolation method used in continual learning. We denote 3) as upper bounds and the others as baselines. Correspondingly, all the systems in 3), 4), 5), 6), 7) and FedSpeech use a pre-trained PWG as the vocoder for a fair comparison. 

The MOS results are shown in Table \ref{table_1}. From the table, we can see that FedSpeech achieves the highest MOS compared with all baselines. It is worth mentioning that FedSpeech outperforms CPG, which illustrates the effectiveness of selectively reusing knowledge from both previous and later speakers. Besides, the results of FedAvg are significantly worse than other methods, which means the gradients from other speakers greatly affect the tone of each speaker. Moreover, the MOS of FedSpeech on VCTK is close to multi-task training (the upper bound). These results demonstrate the advantages of FedSpeech for federated multi-speaker TTS tasks.

\begin{table}
\centering
\begin{tabular}{lrrrrr}  
\toprule
Method & Task 1 & Task 5 & Task 10 & $Avg.$ & $\gamma$ \\
\midrule
Scratch     & 0.8600  & 0.8725 & 0.8647 & 0.8571 & 10   \\
Finetune    & 0.8600  & 0.8782 & 0.8802 & 0.8651 & 10   \\
Multi-task  & 0.8845  & 0.8738 & 0.8784 & 0.8738 & 1.7   \\
FedAvg      & 0.7566  & 0.5736 & 0.7057 & 0.7020 & 1.7   \\
CPG         & 0.8550  & 0.8798 & 0.8847 & 0.8688 & 1.7   \\
\midrule
FedSpeech   & 0.8861  & 0.8884 & 0.8833 &  \textbf{0.8786} & 1.7   \\
\bottomrule
\end{tabular}
\caption{The comparison of speaker similarity between baselines and FedSpeech. $Avg.$ means the average of 10 tasks, and $\gamma$ means the model expansion rate compared with FedSpeech with 256 hidden size.}
\label{table_2}
\end{table}

\subsection{Comparison Experiments for Speaker Similarity in TTS}
\label{spk_sim}
We conduct the speaker similarity evaluation on the test set to measure the similarity between the synthesized audio and the ground truth audio. To exclude other interference factors, we keep the text content consistent among different models. For each task, we derive the high-level representation vector that summarizes the characteristics of the speaker's voice using the encoder\footnote{https://github.com/resemble-ai/Resemblyzer} implemented from Wan \textit{et al.} \shortcite{wan2018generalized}. Specifically, the encoder is a 3-layer LSTM with projection, which is pretrained for extracting speaker's tone embeddings. Cosine similarity is a standard measure of similarity of speaker representation vectors, and is defined as $cos\_sim(A,B) = A \cdot B/ \left \| A \right \| \left \| B \right \| $. The results range from -1 to 1, and higher values indicate that the vectors are more similar. We calculate the cosine similarity between the speaker representation vectors of the synthesized audios and the ground truth audios as the criterion for evaluation. We compare the results of the audio samples generated by our model with those systems described in subsection \ref{audio_quality}.

The results are shown in Table \ref{table_2}. Our FedSpeech scores the highest on average, and even higher than multi-task, the upper bound. It means FedSpeech can retain the voice of each speaker better in the inference stage and demonstrates the effectiveness of parameter isolation. Moreover, in task 1 the result of FedSpeech is significantly higher than CPG. It can be seen that selectively reusing knowledge from both previous and later speakers brings great advantages to the speakers so that all speakers in federated multi-speaker TTS task can obtain better voices.

\subsection{Ablation Studies} We conduct ablation studies to demonstrate the effectiveness of several components in FedSpeech, including the gradual pruning masks and the selective masks. We conduct audio quality and speaker similarity evaluation for these ablation studies. In this experiment, the model is well trained with our proposed first round of training so that we can focus on the effectiveness of the proposed masks.

\paragraph{Audio Quality.} For measuring audio quality, we conduct the MOS evaluation in which each audio is judged by 10 native English speakers. As shown in Table \ref{table_3}, removing the gradual pruning masks or removing the selective masks does not result in significant quality decline, which means the selective masks have the ability to automatically select weights preserved by the gradual pruning masks. However, removing both types of masks leads to catastrophic quality degradation.

\paragraph{Similarity.} We conduct the speaker similarity evaluation as subsection \ref{spk_sim}. As shown in Table \ref{table_3}, simply removing the selective masks or the gradual pruning masks results in slight performance degradation, while removing both of them leads to a catastrophic decline. It can be seen that the gradual pruning masks perfectly preserve the tone of each speaker. Besides, the selective masks have the ability to automatically select weights preserved by the gradual pruning masks and combining them leads to a better result. 

\begin{table}
\centering
\begin{tabular}{lrrrrr}  
\toprule
Setting & MOS & Similarity \\
\midrule
FedSpeech                             & 3.77 $\pm$ 0.052 & 0.8786     \\
\qquad - GPM                          & 3.74 $\pm$ 0.075 & 0.8725    \\
\qquad - SM                            & 3.72 $\pm$ 0.074 & 0.8722     \\
\qquad - SM - GPM                     & 3.23 $\pm$ 0.099 & 0.6304    \\
\bottomrule
\end{tabular}
\caption{MOS and speaker similarity comparison in the ablation studies. SM means the selective masks, GPM means the gradual pruning masks and similarity is the cosine similarity described in subsection \ref{spk_sim}. }
\label{table_3}
\end{table}

\section{Conclusions}
In this work, we have proposed FedSpeech, a high-quality multi-speaker TTS system, to address the ``data island" issue in federated multi-speaker TTS tasks. FedSpeech is implemented based on the two rounds of training with the feed-forward transformer network proposed in Ren \textit{et al.} \shortcite{ren2020fastspeech} and consists of several key components including the selective masks, the progressive pruning masks, and the private speaker module. Experiments on a reduced VCTK dataset (the training set is reduced to a quarter for each speaker to simulate low-resource language scenarios) demonstrate that our proposed FedSpeech can nearly match the upper bound, multi-task training in terms of speech quality, and even significantly outperforms all systems in speaker similarity experiments. For future work, we will continue to improve the quality of the synthesized speech and propose a new mask strategy to compress the model and speed up training. Besides, we will also apply FedSpeech to zero-shot multi-speaker settings by using the private speaker module to generate our proposed masks.

\vspace{1cm}


\bibliographystyle{named}
\bibliography{ijcai21}

\end{document}